\documentclass[10pt]{article}
\usepackage{amsmath,amsfonts}
\usepackage{multirow}
\usepackage{amssymb}
\usepackage[colorlinks=true,urlcolor=blue]{hyperref}
\usepackage{graphicx}
\usepackage{nicefrac}
\usepackage{ulem}
\usepackage{cancel}
\usepackage{multirow}
\hypersetup{                
backref = true,             
pagebackref = true,                 
hyperindex = true,          
colorlinks = true,          
breaklinks = true,          
urlcolor = blue,            
linkcolor = blue,           
bookmarks = true,           
bookmarksopen = true,       
citecolor=red,
}

\newcommand{\bpm}{\begin{pmatrix}}
\newcommand{\epm}{\end{pmatrix}}
\newcommand{\qeed}{\hfill\textrm{QED}\break\null}

\advance \textheight by \topskip
\makeatletter

\def\cc#1{\kern .7em\hfill #1 \hfill\kern .7em}

\newcommand{\rr}{\color{black}}
\newcommand{\bb}{\color{black}}

\newcommand{\beqa}{\begin{eqnarray}}
\newcommand{\eeqa}{\end{eqnarray}}

\newcommand{\nn}{\nonumber}

\newcommand{\ii}{{\mathrm i}}
\newcommand{\g}{{\mathfrak{g}}}
\newcommand{\no}{{}_\circ \hskip -.17truecm {}^\circ \hskip .1 truecm}

\title{Vertex operator for generalised Kac--Moody algebras  associated to the two-sphere and the two-torus}
\begin{document} 
\maketitle
\begin{center}
Rutwig Campoamor-Stursberg$^{1\ast}$, Michel Rausch de Traubenberg$^{2\dagger}$
\medskip

$^1$  Instituto de Matem\'atica Interdisciplinar and Dpto. Geometr\'\i a y Topolog\'\i a, UCM, E-28040 Madrid, Spain\\

$^3$ Universit\'e de Strasbourg, CNRS, IPHC UMR7178, F-67037 Strasbourg Cedex, France\\  

\end{center}
  \noindent $^\ast$ Email: rutwig@ucm.es

\noindent $^\dagger$ Email:  Michel.Rausch@iphc.cnrs.fr 

\medskip

\begin{abstract}
We pursue our study of generalised Kac-Moody and Virasoro algebras defined on compact homogeneous manifolds.
 Extending the well-known Vertex operator in the case of the two-torus
or the two-sphere, we obtain explicit bosonic realisations of the semi-direct product of the extension of Kac-Moody and Virasoro algebras on $\mathbb S^1 \times \mathbb S^1$ and $\mathbb S^2$, respectively. As for the fermionic realisation previously constructed, in order to have well defined algebras, we introduce, beyond the usual normal ordering prescription, a regulator and regularise
infinite sums  by means of  the Riemann $\zeta$-function.

{\bf keywords}: Kac-Moody; Virasoro algebras; bosons realization; regularisation.
\end{abstract}


 It is well known that Kac-Moody \cite{Kac, Kac2, Moo, Mdo,ps,go} and Virasoro \cite{bpz}  algebras are intimately related to the one-dimensional sphere $\mathbb S^1$.   Recently, a class of infinite dimensional Lie
algebras associated to compact real manifolds ${\cal M}$,   with ${\cal M}$ either  a compact
Lie group $G_c$ or  a compact homogeneous space $ G_c/H$ 
 ($H$ being  a closed subgroup of $G_c$)  has been studied \cite{rmm}.
  Extensions of Kac-Moody algebras of this type 
have also  been considered in \cite{bars, KT,Frap} for  ${\cal M} = \mathbb S^2$ and
${\cal M}= \mathbb S^1\times \cdots \times \mathbb S^1$,  
 as well as in \cite{multil, jap} (in the first of
these articles, the algebra associated to $\mathbb S^1\times \cdots \times \mathbb S^1$   is called a multi-loop algebra). 
 The problem of the existence of central extensions, as well as their representation theory in the context of structural properties and representations of either  $G_c$ or ${\cal M}$ were inspected in \cite{rmm}. \
 Following this analysis, it was realised in \cite{rm} that the Kac-Moody algebra and the corresponding Virasora algebras associated to  ${\cal M}=\mathbb S^2$ or ${\cal M}=\mathbb S^1\times \mathbb S^1$ can indeed
 be constructed in a natural manner from the usual Kac-Moody and Virasoro algebras,
  a fact that enables
us  to derive an explicit fermion realisation.

In this note we pursue our study of Kac-Moody and Virasoro algebras associated to
$\mathbb S^2$ and  $\mathbb S^1\times \mathbb S^1$  and derive an explicit boson realisation, which can be  interpreted as a natural extension of the vertex construction \cite{fk, seg}, along the lines of \cite{rm}.\\

We first recall some properties of simply laced Lie algebras  \cite{Mdo}. Let $\mathfrak g$ be a compact simply laced Lie algebra of rank $r$ and let $\Sigma$ be the set of roots of $\g$. Since $\g$ is simply laced, all roots have the same length  and  are normalised  such that their square equals
two. Furthermore,  for any two roots $\alpha,\beta$ such that $\alpha + \beta$ is a root, we have $\alpha\cdot \beta=-1$.
In the Cartan-Weyl basis $\g=\{H^i, E^\alpha, i=1,\dots,r, \alpha\in \Sigma\}$, the Lie brackets take the form

\beqa
\label{eq:g}
\big[H^i, H^j\big]=0 \ , &&
\big[H^i, E_\alpha\big]= \alpha^i E_\alpha \ , \ \ \ 
\big[E_\alpha, E_\beta\big] = \left\{\begin{array}{ll}
\epsilon(\alpha,\beta) E_{\alpha+\beta}&\text{if} \ \ \alpha+ \beta\in \Sigma\\
\alpha\cdot H &\text{if} \ \ \alpha + \beta=0\\
0&\text{otherwhise}
\end{array}
\right.
\eeqa
where $\epsilon(\alpha,\beta)=\pm$.
 Let $\Lambda_R(\g)$ be the root lattice of $\g$. As for all $\alpha,\beta \in \Lambda_R(\g)$, we have $\alpha\cdot\beta\in \mathbb Z$, and the root lattice is said to be an integer lattice. Since $\g$ is compact, the scalar product is Euclidean and the lattice is called Euclidean (with $\alpha\cdot \beta$ being the corresponding scalar product for $\alpha,\beta\in
 \Lambda_R(\g)$).
The key point in the construction below, as shown by Frenkel and Kac \cite{fk} (see also \cite{dG}), is to extend this sign function
$\epsilon(\alpha,\beta)$  for any $\alpha, \beta\in \Lambda_R(\g)$.  This
extension satisfies 
\beqa
\label{eq:eps}
\epsilon(\alpha,\beta) =(-1)^{\alpha\cdot \beta+ \alpha\cdot \alpha  \;\beta\cdot \beta} \epsilon(\beta,\alpha)\ , \ \ \ \ 
\epsilon(\alpha,\beta) \epsilon(\alpha+\beta,\gamma)=\epsilon(\alpha,\beta+\gamma) \epsilon(\beta,\gamma) \ ,
\eeqa
 for any $\alpha,\beta \in \Lambda_R(\g)$.
A lattice  satisfying these requirements is called an $\epsilon-$lattice. In fact, for any simply laced Lie algebra $\g$, the root lattice $\Lambda_R(\g)$ is an $\epsilon-$lattice (see {\it e.g.} \cite{go-l} for details).  The $\epsilon-$function is not uniquely defined,
 and it admits  a kind of $\mathbb Z_2-$gauge invariance, as the new function
\beqa
\epsilon'(\alpha,\beta)= \eta_\alpha \eta_\beta \eta_{\alpha + \beta} \epsilon(\alpha,\beta) \ , \nn
\eeqa
satisfies  condition (\ref{eq:eps}),
where $\eta_\alpha, \eta_\beta, \eta_{\alpha + \beta} =\pm 1$. Fixing the gauge enables to chose  an $\epsilon-$function such that
\beqa
\label{eq:eps2}
\epsilon(0,\alpha) = \epsilon(\alpha,0) =\epsilon(\alpha,-\alpha)=1 \ \ \ \ 
\epsilon(\alpha,\beta)=\epsilon(-\beta,-\alpha) = \epsilon(-\alpha, \alpha +\beta)\ , 
\eeqa
 Details can be found for instance in \cite{go}.\\

Now we turn to the boson realisation of the Kac-Moody and Virasoro algebras associated to the two-sphere and the two-torus.  We first introduce $r=\text{rk}(\g)$ scalar fields $Q^i$ and their associated momenta $P^i$. Instead of parameterising the two-torus (resp. the two-sphere) by the  angular variables $(\varphi_1,\varphi_2) \in [0,2\pi[\times[0,2\pi[$ (resp. $(\varphi,u) \in [0,2\pi[\times[-1,1]$ ---with $u=\cos \theta$), we  consider in our two-dimensional manifolds the local coordinates  $(z,\varphi)$ (resp. $(z,u)$) with $z \in \mathbb C$,
as done in \cite{rm},
 in order to use techniques of Conformal Field Theory. With these notations, we introduce  for the two-torus $\mathbb{S}^1\times \mathbb{S}^1$

\beqa
\label{eq:QT2}
Q^i(z,\varphi)&=&  \sum \limits_{n \in \mathbb Z}\Bigg[  q_n^i  -\ii  p_n^i \ln z
+ \ii \sum \limits_{m>0} \Big(\frac 1 m \alpha^i_{m,n} z^{-m} -\frac 1 m \alpha^i_{-m,n} z^m\Big)
\Bigg]e^{-\ii n\varphi} 
 \ ,\nn \\
 &=& q^i(\varphi)-\ii p^i(\varphi) \ln z + \ii \sum \limits_{m>0}\frac 1 m \Big(\alpha^i_m(\varphi) z^{-m} - \alpha^i_{-m}(\varphi) z^m\Big) \\ 
P^i(z,\varphi)&=& \ii z \frac{\text{d} Q^i(z,\varphi)}{\text{d} z}=
\sum \limits_{n\in \mathbb Z} \Bigg[ p_n^i + \sum \limits_{m>0} \Big( \alpha^i_{m,n} z^{-m} +\alpha^i_{-m,n} z^m\Big)\Bigg] e^{-\ii n \varphi} \ ,
   \nn\\
  &=& p^i(\varphi) + \sum \limits_{m>0}\Big(\alpha^i_m(\varphi) z^{-m} + \alpha^i_{-m}(\varphi) z^m\Big)\nn  
\eeqa
 and for the two-sphere $\mathbb{S}^2$
\beqa
\label{eq:QS2}
Q^i(z,u)&=&  \sum \limits_{\ell \ge 0} \Big(q_\ell^i -\ii p_\ell^i\ln z\Big)Q_{\ell 0}(u)
+\ii  \sum \limits_{m>0}  \sum \limits_{\ell \ge m} \frac 1 m\Big(\alpha_{\ell m}^i z^{-m}- \alpha_{\ell -m}^i z^{m}\Big)Q_{\ell,m}(u)\nn  \\
&=& q^i(u) -\ii p^i(u) \ln z + \ii \sum \limits_{m>0}\frac 1 m \Big(\alpha^i_m(u) z^{-m} - \alpha^i_{-m}(u) z^m\Big) \\ 
P^i(z,u)&=&   \ii z \frac{\text{d} Q^i(z,u)}{\text{d} z}=\sum \limits_{\ell \ge 0}  p_\ell^i Q_{\ell 0}(u)
+ \sum \limits_{m>0}   \sum \limits_{\ell \ge m} \Big(\alpha_{\ell m}^i z^{-m} +\alpha_{\ell -m}^i z^{m}\Big)Q_{\ell,m}(u)\ , \nn\\
&=&p^i(u)  + \sum \limits_{m>0}\Big(\alpha^i_m(u) z^{-m} + \alpha^i_{-m}(u) z^m\Big)\nn  
\eeqa
where $Q_{\ell m}$ are the normalised associated Legendre functions (see \cite{rm} for notations), and
\beqa
\label{eq:Bn}
{\cal B}_m=\Big\{Q_{\ell m}, \ell\ge |m|\Big\}\ , \ \ \forall  m \in \mathbb Z \ ,
\eeqa
constitutes a Hilbert basis for the square integrable functions on $[-1,1]$. We observe that the second and fourth lines in \eqref{eq:QT2} and \eqref{eq:QS2} correspond to the usual mode expansions considered in  the context of strings. In other words, we assume in our construction that the string modes ($\alpha_m^i, p^i$ and $q^i$) have mode expansions in the $\varphi-$direction for the two-torus, and in the
$u-$direction for the two-sphere. This is the key observation to obtain, from the usual Virasoro and
Kac-Moody algebras,  a corresponding analogue on $\mathbb S^1\times \mathbb S^1$ and  $\mathbb S^2$. This mode expansion has also
been considered in \cite{rm} to construct fermion realisations of generalised Kac-Moody and Virasoro algebras. Note finally that, in the case of the two-torus, the construction is dissymmetric in  both directions. This is in fact a consequence of our ordering prescription (see \eqref{eq:ord} below), but is in accordance with unitarity of representations of generalised Kac-Moody algebras, as shown in \cite{rmm}.

\medskip
We assume the quantisation relations
\beqa
\label{eq:quant}
\begin{array}{cllclllc}
\big[\alpha^i_{m,p}, \alpha^j_{n,q}\big]&=& m \delta^{ij} \delta_{m+n}\delta_{p+q} \ ,& \big[p_m^i,q_n^j]
&=& -\ii \delta^{ij}
\delta_{m+n} &\text{for}&\mathbb S^1\times \mathbb S^1\\[0.25cm]
\big[\alpha^i_{\ell,m}, \alpha^j_{\ell',m'}\big]&=& m \delta^{ij} \delta_{\ell \ell'}\delta_{m+m'} \ , & \big[p_\ell^i,q_{\ell'}^j]& =& -\ii \delta^{ij}
\delta_{\ell \ell'} &\text{for}&\mathbb S^2
\end{array}
\eeqa
 as well as the reality conditions
\beqa
\label{eq:real}
\begin{array}{llllc}
(\alpha^i_{m,p})^\dag= \alpha^i_{-m,-p}\ , & (p_m^i)^\dag =p_{-m}^i\ , &  (q_m^i)^\dag =q_{-m}^i &\text{for}&\mathbb S^1\times S^1\\[0.25cm]
(\alpha^i_{\ell,m})^\dag =\alpha^i_{\ell,-m} \ ,& (p_\ell^i)^\dag = p_\ell^i\ ,&(q_{\ell}^j)^\dag = q_{\ell}^j &\text{for}&\mathbb S^2\ .
\end{array}
\eeqa
 Note that in equations \eqref{eq:quant} and \eqref{eq:real} we don't have the $(-1)^m$ coefficients for the two-sphere, as it is the case for
fermions in \cite{rm}. This is simply due to the fact that in the expansion \eqref{eq:QS2}, we have substituted  $Q_{\ell,m}(u)=(-1)^m Q_{\ell,-m}(u)$
by  $(-1)^m Q_{\ell,-m}(u)$ for negative values of $m$, thus reabsorbing the sign factor. Defining the order relation
\beqa
\label{eq:ord}
\begin{array}{llllc}
(m,p) > 0& \Longleftrightarrow &m >0 &\text{for}&\mathbb S^1\times \mathbb S^1\\
(\ell,m)> 0&\Longleftrightarrow& m >0 &\text{for}&\mathbb S^2
\end{array}
\eeqa
the annihilation operators are $\alpha^i_{m,p}, m>0,  p_p^i, \forall p\in \mathbb Z$ for the two-torus and
 $\alpha^i_{\ell,m},  m>0, \ell \ge m,  p_\ell^i,  \ell \ge 0$ for the two-sphere.
 As usual in Quantum Field Theory, to avoid divergences we define the normal ordering prescription
\beqa
 \label{eq:no}
 \begin{array}{lllllllc}
\no \alpha_{m,p}^i \alpha^j_{n,q}\no &=& \left\{
\begin{array}{lll}
 \alpha^j_{n,q}\alpha_{m,p}^i & m>0& \forall q\in\mathbb Z\\
 \alpha_{m,p}^i\alpha^j_{n,q} &  m<0& \forall q\in\mathbb Z
 \end{array}\right.&
 \no p^i_m q^j_n\no&=&q^j_n p^i_m&\text{for}& \mathbb S^1 \times \mathbb S^1\\[0.35cm]
 \no \alpha_{\ell,m}^i \alpha^j_{\ell',m'}\no &=& \left\{
\begin{array}{lll}
 \alpha^j_{\ell', m'}\alpha_{\ell, m}^i & m>0&  \ell \ge |m| \\
 \alpha_{\ell, m}^i\alpha^j_{\ell',m'} &  m<0& \ell \ge |m|
 \end{array}\right.&
 \no p^i_\ell q^j_{\ell'}\no&=&q^j_{\ell'} p^i_\ell&\text{for}& \mathbb S^2\ .
 \end{array}
\eeqa
However, as  already observed in \cite{rm}, the normal ordering prescription \eqref{eq:no} is  still not enough to avoid divergences.
To circumvent this difficulty and obtain an appropriate regularisation procedure, we define the $\epsilon-$regularised bosons
(with $\epsilon>0$)
\beqa
\label{eq:bosreg}
\begin{array}{lllll}
Q^i_\epsilon(z,\varphi)&=&  \sum \limits_{n \in \mathbb Z}\Bigg[  q_n^i  -\ii  p_n^i \ln z
+ \ii \sum \limits_{m>0} \Big(\frac 1 m \alpha^i_{m,n} z^{-m} -\frac 1 m \alpha^i_{-m,n} z^m\Big)
\Bigg]e^{-\ii n\varphi} e^{-\epsilon(|n| -\frac12)} \\&&\text{for} \ \ \mathbb S^1 \times \mathbb S^1\\
Q^i_\epsilon(z,u)&=&  \sum \limits_{\ell \ge 0} \Big(q_\ell^i -\ii p_\ell^i\ln z\Big)Q_{\epsilon,\ell 0}(u)
+\ii  \sum \limits_{m>0}  \sum \limits_{\ell \ge m} \frac 1 m\Big(\alpha_{\ell m}^i z^{-m}- \alpha_{\ell -m}^i z^{m}\Big)Q_{\epsilon,\ell m}(u)\\
&&\text{for}\mathbb S^2
\end{array}
\eeqa
with $Q_{\epsilon,\ell m}$ the $\epsilon-$regularised associated Legendre functions \cite{rm}.  A completely analogous
expression is obtained for the  $P_\epsilon^i$s.
Associated to these $\epsilon-$regularised bosons we define the functions
\beqa
\label{eq:deleps}
\begin{array}{llc}
\delta_\epsilon(\theta-\varphi)  = \sum \limits_{m \in \mathbb  Z } e^{-\ii m (\theta-\varphi)} e^{-2 \epsilon(|m|-\frac12)} & \text{for}&\mathbb S^1 \times \mathbb S^1\\[0.45cm]
\delta^m_\epsilon(u-v) = \sum \limits_{\ell\ge |m|} Q_{\epsilon,\ell,m}(u)  Q_{\epsilon,\ell,m}(v)  & \text{for}&\mathbb S^2\ .
\end{array}
\eeqa
When $\epsilon \to 0$, on the one hand the regularised bosons \eqref{eq:bosreg} reduce to  either \eqref{eq:QT2} or \eqref{eq:QS2}, while on the other hand, the functions \eqref{eq:deleps} reduce to the usual $\delta-$distributions on either $\mathbb S^1\times \mathbb S^1$ or $\mathbb S^2$ (in the second case, the $\delta-$distribution is associated to the basis \eqref{eq:Bn}). However, as the generators of
the Kac-Moody and the Virasoro algebras on the two-torus and the two-sphere are bilinear in the bosons (see below), the Wick theorem involves a double contraction, and in particular a term  of the type $\delta_\epsilon(\varphi-\theta)^2$ in the case of the two-torus. This term is perfectly defined whilst the limit for $\epsilon \to 0$ is not defined.
This situation has already been encountered in \cite{rm}, when constructing a fermion realisation. We recall here the
salient points of our regularisation procedure. In particular, to  obtain a well-defined limit, we propos{e {\bb regularisation} prescription. This prescription is done} {\rr in two steps}. In the first step, we substitute
(in the case of the two-torus) $\delta_\epsilon(\theta-\varphi)^2$ by  $\delta_{\epsilon_1} (\theta-\varphi)\delta_{\epsilon_2}(\theta-\varphi)$ and take the limit $\epsilon_1\to 0, \epsilon_2\to 0$  separately. This leads to 
\beqa
\lim_{\epsilon_1 \to 0} \delta_{\epsilon_1} (\theta-\varphi)\delta_{\epsilon_2}(\theta-\varphi)=  \delta (\theta-\varphi)\delta_{\epsilon_2}(\theta-\varphi)=  \delta (\theta-\varphi)\delta_{\epsilon_2}(0) \ \nn
\eeqa
with $\delta_{\epsilon_2}(0)$ finite. In a second step, the divergent limit $\epsilon_2\to 0$  for $\delta_{\epsilon_2}(0)$ is regularised using the $\zeta-$regularisation (where $\zeta$ denotes  the Riemann $\zeta-$function) \cite{rm}.
A similar regularisation process  is valid for the  {\rr two-sphere}. In both cases, we introduce the regulated expressions~\cite{rm}
\beqa
\begin{array}{llc}
\delta_{\text{reg}}(0)  =\lim\limits_{\epsilon\to 0} \delta_\epsilon(0)|_{\text{reg}} =1 &\text{for}&\mathbb S^1 \times \mathbb S^1\\[0.25cm]
\delta^m_{\text{reg}}(0)  =\lim\limits_{\epsilon\to 0} \delta^m_\epsilon(0)|_{\text{reg}} =1 &\text{for}&\mathbb S^2\ .
\end{array}
\eeqa
Using the normal ordering prescription \eqref{eq:no}, the quantisation relation \eqref{eq:quant}  and applying the Wick
theorem (omitting the regular term  --- analogously, in all expressions below we shall  omit  the regular term), we obtain
\beqa
\label{eq:QQT2}
\begin{array}{lllll}
Q_\epsilon^i(z,\theta) Q_\epsilon^i(w,\varphi)&=& -\delta^{ij} \ln(z-w) \delta_\epsilon(\theta-\varphi) \ , &|z|> |w| \\[0.5ex]
P_\epsilon^i(z,\theta) Q_\epsilon^i(w,\varphi)&=& \displaystyle - \frac{\ii \delta^{ij} z}{z-w} \delta_\epsilon(\theta-\varphi) \ , & |z|> |w| \\[0.7ex]
P_\epsilon^i(z,\theta) P_\epsilon^i(w,\varphi)&=& \displaystyle \frac{\delta^{ij} zw}{(z-w)^2} \delta_\epsilon(\theta-\varphi) \ , &|z|> |w|
\end{array}
\eeqa
for the two-torus,  while for the two-sphere we have
\beqa
\label{eq:QQS2}
\begin{array}{lllll}
Q_\epsilon^i(z,u) Q_\epsilon^i(w,v)&=& \delta^{ij} \big(-\ln z \delta_\epsilon^0(u-v)
+ \sum \limits_{m>0} \sum \limits_{\ell \ge m} \frac 1 m\Big(\frac w z \Big)^m \delta_\epsilon^m(u-v)\\
&\equiv& -\delta^{ij}
L_\epsilon(z,w,u,v)\ ,& |z|> |w|\\[0.6ex]
P_\epsilon^i(z,u) Q_\epsilon^i(w,v)&=& - \ii \delta^{ij} \sum \limits_{m\ge 0}
\Big(\frac w z \Big)^m \delta_\epsilon^m(u-v) \\
&\equiv& -\ii \delta^{ij} \Delta_\epsilon(z,w,u.v)\ ,&  |z|> |w| \\[0.6ex]
P_\epsilon^i(z,u) P_\epsilon^i(w,v)&=& \delta^{ij} \sum \limits_{m\ge 0} m 
\Big(\frac w z \Big)^m \delta_\epsilon^m(u-v)  . & |z|> |w|
\end{array}
\eeqa
 In the limit $\epsilon\to 0$,  the equations \eqref{eq:QQT2} and \eqref{eq:QQS2} are not singular, and in both cases we obtain
\beqa
\label{eq:QQreg}
\begin{array}{lllll}
Q^i(z,x) Q^i(w,y)&=& -\delta^{ij} \ln(z-w) \delta(x-y) \ , &|z|> |w| \\[0.7ex]
P^i(z,x) Q^i(w,y)&=& \displaystyle - \frac{\ii \delta^{ij} z}{z-w} \delta(x-y) \ , & |z|> |w| \\[0.7ex]
P^i(z,x) P^i(w,y)&=& \displaystyle  \frac{\delta^{ij} zw}{(z-w)^2} \delta(x-y) \ , &|z|> |w|
\end{array}
\eeqa
 where $(x,y)=(\varphi,\theta)$ for $\mathbb S^1\times \mathbb S^1$ and  $(x,y)=(u,v)$ for $\mathbb S^2$. 
The Hilbert space associated to the bosons \eqref{eq:QT2} and \eqref{eq:QS2} is built from the vacuum  state $\big|0\big>$ defined by
\beqa
\label{eq:vac}
\begin{array}{llllllllc}
\alpha_{m,n}^i \big|0\big> &=& 0\ ,\  \ m>0 \ ,\ &  p_n^i \big|0\big> &=&0\ , &\forall n \in \mathbb Z & \text{for}& 
\mathbb S^1\times S^1 \\[0.5ex]
\alpha_{\ell, m}^i \big|0\big> &=& 0\ ,\ \  m>0 \ , & p_\ell^i \big|0\big> &=&0 \ ,  & \forall \ell  \ge 0 & \text{for} &
\mathbb S^2
\end{array}
\eeqa
Introduce now 
$\lambda \in \Lambda_W(\g)$ the weight lattice of $\g$ ($\Lambda_\Sigma(\g) \subset \Lambda_W(\g))$ and   define 
\beqa
\label{eq:wave}
\begin{array}{lllllc}
\big|\lambda,\varphi\big> &=& e^{\ii \lambda\cdot q(\varphi)}\big|0\big>  \ , & q^i(\varphi)=
\sum\limits_{n\in \mathbb Z} q_n^i e^{-\ii n \varphi} &
\text{for} &\mathbb S^1 \times \mathbb S^1 \\
\big|\lambda,u\big> &=& e^{\ii \lambda\cdot q(u)}\big|0\big>  \ , &   q^i(u)=
\sum\limits_{\ell \in \mathbb N} q_\ell^i Q_{\ell,0}(u)&
\text{for} & \mathbb S^2
\end{array}
\eeqa
which  are normalised as follows
\beqa
\big<\lambda,\varphi\big|\mu,\theta\big> &=& \delta_{\mu \nu} \delta(\varphi -\theta) \ \ \text{for} \ \ \mathbb S^1\times\mathbb
S^1\nn\\
\big<\lambda,u\big|\mu,v\big> &=& \delta_{\mu \nu} \delta(u-v) \ \ \text{for} \ \ \ \ \  \mathbb S^2\ . \nn
\eeqa
We introduce the vertex operator
\beqa
\label{eq:Ver}
U_\epsilon^\alpha(z,x)= z^{\alpha\cdot\alpha/2} \; \no e^{\ii \alpha \cdot Q_\epsilon(z,x)}\no
\eeqa
with $x=\varphi$ for $\mathbb S^1\times \mathbb S^1$ and  $x=u$ for $\mathbb S^2$. The limit $\epsilon\to0$ of \eqref{eq:Ver}
is well defined.  This procedure constitutes a direct generalisation of the Vertex operator considered in string theory \cite{fu1,fu2,ger,go}. Acting on the states \eqref{eq:wave}, the vertex operator
is single valued, as $\lambda$ belongs to the weight lattice (see {\it e.g.} \cite{go}).
Again, using the Wick theorem, we obtain
\beqa
\begin{array}{llllc}
P_\epsilon^i(z,\varphi) U_\epsilon^\alpha(w,\theta)&=&\alpha^i\frac{ z}{z-w} \delta_\epsilon(\varphi-\theta) U_\epsilon^\alpha(w,\theta)&
\text{for}&\mathbb S^1 \times \mathbb S ^1\\[0.6ex]
P_\epsilon^i(z,u) U_\epsilon^\alpha(w,v)&=&\alpha^i\Delta_\epsilon(z,w,u,v) U_\epsilon^\alpha(w,v)&
\text{for}&\mathbb S^2 
\end{array} \ \ |z|>|w|\ .
\eeqa
 In the limit $\epsilon \to 0$, both relations above are regular and lead to
\beqa
\label{eq:PU}
P^i(z,x) U^\alpha(w,y)&=&\alpha^i\frac{ z}{z-w} \delta(x-y) U^\alpha(w,y) \ , \ \ |z| >|w|
\eeqa
where $(x,y)=(\theta,\varphi)$ for $\mathbb S^1 \times \mathbb S^1$ and $(x,y)=(u,v)$ for $\mathbb S^2$.
Finally,  pushing the creation operators of  $U_\epsilon^\beta$  to the left of the annihilation operators of
$U_\epsilon^\alpha$, using \eqref{eq:no}, we obtain (for  $|z| >|w|$)
\beqa
\begin{array}{llllc}
U_\epsilon^\alpha(z,\theta) U_\epsilon^\beta(w,\varphi)&=& \no (z-w)^{\alpha\cdot \beta \delta_\epsilon(\theta-\varphi)}
 e^{ \ii \alpha \cdot Q_\epsilon(z,\theta) + \ii \beta\cdot Q_\epsilon(w,\varphi)}\no&\text{for}&\mathbb S^1 \times \mathbb S^1\\[0.26cm]
U_\epsilon^\alpha(z,u) U_\epsilon^\beta(w,u)&=& \no e^{\alpha\cdot \beta L_\epsilon(z,w,u,v)}
 e^{ \ii \alpha \cdot Q_\epsilon(z,\theta) + \ii \beta\cdot Q_\epsilon(w,\varphi)}\no&\text{for}&\mathbb S^2\\
\end{array}\nn
\eeqa
It is known that, in the construction of the usual Kac-Moody algebras, the Vertex operator has to be corrected by a two-cocycle
(see {\it e.g.} \cite{go} for the case of usual Kac-Moody algebras). Analogously, we define   (associated to the vacuum specified by the momenta $\lambda \in \Lambda_W(\g)$, see \eqref{eq:wave})
\beqa
c_\alpha(x) =\sum\limits_{\beta \in \Lambda_R(\g)} \epsilon(\alpha,\beta) \big|\beta + \lambda, x\big>\big<\beta + \lambda,x\big| \nn
\eeqa
with $x=\varphi$ for $\mathbb S^1\times \mathbb S^1$ and $x=u$ for $\mathbb S^2$.
Next we introduce 
\beqa
\label{eq:chat}
\hat c_\alpha(x) = e^{\ii  q(x)\cdot \alpha} c_\alpha(x) =
\sum\limits_{\beta \in \Lambda_R(\g)} \epsilon(\alpha,\beta) \big|\alpha+\beta + \lambda,x\big>\big<\beta + \lambda,x\big| 
\eeqa
where $q^i(\varphi)$  and $q^i(u)$ are defined in  \eqref{eq:wave}.
A direct computation  using \eqref{eq:eps} shows that 
\beqa
\label{eq:c}
\hat c_\alpha(x) \hat c_\beta(y) = (-1)^{\alpha\cdot\beta + \alpha\cdot \alpha \;\beta\cdot \beta}\hat c_\beta(y) \hat c_\alpha(x) =
\epsilon(\alpha,\beta)\hat
c_{\alpha + \beta} (y) \delta(x-y)\ 
\eeqa
and
\beqa
\hat c_\alpha^\dag(x)= \hat c_{-\alpha}(x) \ . \nn
\eeqa
If we now define $\hat U_\epsilon^\alpha(z,x) = U_\epsilon^\alpha(z,x)  c_\alpha(x)$ and regularise
$\hat U_\epsilon^\alpha(z,x) \hat U_\epsilon^\beta(w,y)$, we obtain for  the two-torus (for $z| >|w|$)
\beqa
\label{eq:UU}
\hat U_\epsilon^\alpha(z,\theta)\hat U_\epsilon^\beta(w,\varphi) &=&\epsilon(\alpha,\beta)    \no (z-w)^{\alpha\cdot \beta \delta_\epsilon(\theta-\varphi)}
e^{ \ii \alpha \cdot Q_\epsilon(z,\theta) + \ii \beta\cdot Q_\epsilon(w,\varphi)}\no\nn\\
&&\hat c_{\alpha +\beta} (\varphi) \;\delta(\theta-\varphi)\nn\\
 &=&\epsilon(\alpha,\beta)    \no (z-w)^{\alpha\cdot \beta \delta_\epsilon(0)}
e^{ \ii \alpha \cdot Q_\epsilon(z,\theta) + \ii \beta\cdot Q_\epsilon(w,\varphi)}\no \nn\\
&&\hat c_{\alpha +\beta} (\varphi) \;\delta(\theta-\varphi)\nn\\
 &\downarrow{\text{reg}}&\\
\hat U^\alpha(z,\theta)\hat U^\beta(w,\varphi)   & =& \epsilon(\alpha,\beta) (z-w)^{\alpha\cdot \beta}\; \no
 e^{ \ii \alpha \cdot Q(z,\theta) + \ii \beta\cdot Q(w,\varphi)}\no \hat c_{\alpha +\beta} (\varphi) \delta(\theta-\varphi)\ ,\nn
\eeqa
 For the two-sphere, a similar procedure leads to (for $z| >|w|$)
\beqa
\label{eq:UUS2}
\hat U^\alpha(z,u)\hat U^\beta(z,v) = \epsilon(\alpha,\beta)(z-w)^{\alpha\cdot\beta} \; \no e^{\ii \alpha\cdot Q(z,x) + \ii \beta\cdot Q(w,y)}\no \hat
c_{\alpha +\beta}(v) \delta(u-v) \ .  
\eeqa
Assuming the mode expansion   (here we change slightly the conventions with respect to \eqref{eq:QS2} since the expansion for $\mathbb S^2$ involves  $Q_{\ell,m}$ and not  $Q_{\ell,{-m}}$  for negative values of $m$)
\beqa
\begin{array}{cllc}
\left.\begin{array}{lll}
P^i(z,\varphi) &=& \sum \limits_{m \in \mathbb Z} \sum \limits_{p \in \mathbb Z} H^i_{m,p} \; z^{-m} \; e^{-\ii p \varphi}\\
\hat U^\alpha(z,\varphi)&=&   \sum \limits_{m \in \mathbb Z} \sum \limits_{p \in \mathbb Z} E_{\alpha,m,p}\;  z^{-m} \; e^{-\ii p \varphi}
\end{array}\hskip .55truecm \right\}&\text{for}&\mathbb S^1 \times \mathbb S^1\\
& &\\
\left.\begin{array}{lll}
P^i(z,u) &=& \sum \limits_{m \in \mathbb Z} \sum \limits_{\ell\ge |m|} H^i_{m,p} \; z^{-m}\;  Q_{\ell m}(u)\\
\hat U^\alpha(z,u)&=& \sum \limits_{m \in \mathbb Z} \sum \limits_{\ell\ge |m|}   E_{\alpha,\ell,m} \;  z^{-m}\;  Q_{\ell m}(u) 
\end{array}\right\}&\text{for}&\mathbb S^2
\end{array}
\eeqa
 and extracting the modes
\beqa
\begin{array}{clc}
\left.\begin{array}{lll}
 H^i_{m,p} &=& \displaystyle\oint \frac{\text{d} z\; z^m}{2 \ii \pi z} \int \limits_0^{2 \pi} \frac{\text{d}\varphi  }{2  \pi }  e^{\ii p \varphi} P^i(z,\varphi)
\\
 E_{\alpha,m,p} &=&   \displaystyle\oint \frac{\text{d} z\; z^m}{2 \ii \pi z} \int \limits_0^{2 \pi} \frac{\text{d}\varphi  }{2  \pi } e^{\ii p \varphi} \hat U^\alpha(z,\varphi)
\end{array}\hskip .5truecm \right\}&\text{for}&\mathbb S^1 \times \mathbb S^1\\
\left.\begin{array}{lll}
H^i_{\ell,m}  &=& \displaystyle \oint \frac{\text{d} z\; z^m}{2 \ii \pi z} \int \limits_{-1}^{1} \frac{\text{d}u  }{2 } P^i(z,u)  Q_{\ell m}(u)\\
 E_{\alpha,\ell,m} &=&\displaystyle \oint \frac{\text{d} z\; z^m}{2 \ii \pi z} \int \limits_{-1}^{1} \frac{\text{d}u  }{2 } \hat U^\alpha(z,u)   Q_{\ell m}(u)
\end{array}\right\}&\text{for}&\mathbb S^2
\end{array}
\eeqa
 Applying  standard techniques of Conformal Field Theory, {\it i.e.},\ integration
in the complex $z-$ and $w-$planes with adapted contours of integration  (see {\it e.g.} \cite{go, dms}) with the  $\varphi$ and  $\theta$  or the $u$ and $v$ integrations being trivial, the equations \eqref{eq:QQreg}, \eqref{eq:PU}, \eqref{eq:UU}
and \eqref{eq:UUS2} lead to
\beqa
\label{eq:KM-S1xS1}
\big[H^i_{m_1,m_2}, H^j_{n_1,n_2}\big]&=& m_1 \delta^{ij} \delta_{m_1 + n_1} \delta_{m_2 + n_2} \ , \nn\\
\big[H^i_{m_1,m_2}, E_{\alpha,n_1,n_2}\big]&=& \alpha^i E_{\alpha,m_1+ n_1,m_2+n_2} \ , \\
\big[E_{\alpha,m_1,m_2}, E_{\beta,n_1,n_2}\big] &=& \left\{\begin{array}{ll}
\epsilon(\alpha,\beta) E_{\alpha+\beta, m_1+n_1,m_2+n_2}&\text{if} \ \ \alpha+ \beta\in \Sigma\\
\alpha\cdot H_{m_1+n_1,m_2+n_2} + m_1 \delta_{m_1 + n_1} \delta_{m_2 + n_2}   &\text{if} \ \ \alpha + \beta=0\\
0&\text{otherwise}
\end{array}
\right.\nn
\eeqa
for the two-torus, and to
\beqa
\label{eq:KM-S2}
\big[H^i_{\ell_1,m_1}, H^j_{\ell_2,m_2}\big]&=& (-1)^{m_1} m_1 \delta^{ij} \delta_{\ell_1,\ell_2} \delta_{m_1 + m_2}   \ , \nn\\
\big[H^i_{\ell_1,m_1}, E_{\alpha,\ell_2, m_2 }\big]&=& c_{\ell_1,m_1,\ell_2,m_2}{}^{\ell_3,m_3} \alpha^i E_{\alpha,\ell_3, m_3} \ , \\
\big[E_{\alpha,\ell_1,m_1}, E_{\beta,\ell_2,m_2}\big] &=& \left\{\begin{array}{ll}
\epsilon(\alpha,\beta) c_{\ell_1,m_1,\ell_2,m_2}{}^{\ell_3,m_3}   E_{\alpha+\beta, \ell_3,m_3}\\
\hskip 1.truecm \text{if} \ \ \alpha+ \beta\in \Sigma\\
c_{\ell_1,m_1,\ell_2,m_2}{}^{\ell_3,m_3}  \alpha\cdot H_{\ell_3,m_3} +  (-1)^{m_1} m_1 \delta_{\ell_1,\ell_2} \delta_{m_1 + m_2}   \\
\hskip 1.truecm \text{if} \ \ \alpha + \beta=0\\
0  \hskip .8truecm
\text{otherwhise}
\end{array}
\right.\nn
\eeqa
 for the two-sphere, where the coefficients $c_{\ell_1,m_1,\ell_2,m_2}{}^{\ell_3,m_3}$
(related to the Clebsch-Gordan coefficients) are given in \cite{rmm}.
We have thus obtained an explicit bosonic realisation of the Kac-Moody algebras associated to the two-torus or the two-sphere with a central charge $k=1$ defined  in \cite{rmm}.  We emphasise that, as stated previously, the Kac-Moody algebra associated to the two-torus is  dissymmetric in both directions. This is a direct consequence of our construction procedure (see the ordering relation \eqref{eq:no}).\\

We now  introduce the generators  of the Virasoro  algebra associated to the free bosons on the two-sphere and the two-torus
\beqa
\label{eq:T}
\begin{array}{llc}
T_\epsilon(z,\varphi) = \frac 12 \sum\limits_{i=1}^r \no P_\epsilon^i(z,\varphi) P_\epsilon^i(z,\varphi) \no &
\text{for}&\mathbb S^1 \times \mathbb S^1\\
T_\epsilon(z,u) = \frac 12 \sum\limits_{i=1}^r \no P_\epsilon^i(z,u) P_\epsilon^i(z,u) \no &
\text{for}&\mathbb S^2
\end{array}
\eeqa
and define the mode expansion, when $\epsilon \to 0$,  by
\beqa
\label{eq:Tmod}
\begin{array}{llc}
T(z,\varphi)  =
\sum\limits _{n\in \mathbb Z} \sum\limits_{p\in \mathbb Z} L_{m,p} z^{-n} e^{-\ii p\varphi}&
\text{for}&\mathbb S^1 \times \mathbb S^1\\[0.25cm]
T(z,u) = \sum\limits _{m\in \mathbb Z} \sum\limits_{\ell\ge |m|} L_{\ell,m} z^{-m}
Q_{\ell m}(u)&
\text{for}&\mathbb S^2
\end{array}
\eeqa
Computing  the products $T_\epsilon T_\epsilon$ and using the Wick theorem, we
proceed to  consider all possible contractions,  where only the double contractions lead to divergences.  The latter
are regularised by the procedure described above.  After regularisation, performing all possible
contractions with omission of the regular part, we obtain 
\beqa  
T(z,\varphi) T(w,\theta)&=&   \frac r 2 \frac{ (zw)^2}{(z-w)^4} \delta(\varphi-\theta)
+ \frac{zw}{(z-w)^2} 2 T(w,\varphi) \delta(\varphi-\theta)\nn\\
&&+ \frac{z}{z-w} w\partial_w T(w,\varphi) \delta(\varphi-\theta)
\  , \ \ |z| > |w| \nn
\eeqa
for the two-torus, and
\beqa  
T(z,u) T(w,v)&=&   \frac r 2 \frac{ (zw)^2}{(z-w)^4} \delta(u-v)
+ \frac{zw}{(z-w)^2} 2 T(w,v) \delta(u-v) \nn\\&&+ \frac{z}{z-w} w\partial_w T(w,v) \delta(u-v)
\  , \ \ |z| > |w| \nn
\eeqa
for the two-sphere. Again using adapted contours of integration in the $z-$ and $w-$ plane (see {\it e.g.} \cite{go, dms}),
the mode expansion \eqref{eq:T} leads to 
\beqa
\label{eq:LL}
\begin{array}{lll}
\big[L_{m,p},L_{n,q}\big]&=&(m-n) L_{n+m, p+q} + \frac{r}{12} m(m^2-1) \delta_{m+n} \delta_{p+q} \\[0.5ex]
&&\hskip 2.truecm \text{for}  \ \ \mathbb S^1 \times \mathbb S^1\\[1.75ex]
\big[L_{\ell_1,m_1},L_{\ell_2,m_ 2}\big]&=&(m_1-m_2) c_{\ell_1,m_1,\ell_2,m_2}{}^{\ell_3,m_1 +m_2} L_{\ell_3, m_1+m_2} \nn\\
&&+(-1)^{m_1} \frac{r}{12} m_1(m_1^2-1)
\delta_{\ell_1 \ell_2} \delta_{m_1 + m_2} \hskip 2.truecm \text{for}  \ \ \ \ \  \mathbb S^2\ . 
\end{array}
\eeqa
 Thus we have obtained  an explicit bosonic realisation of the Virasoro algebras associated to the two-torus or the two-sphere with a central charge $c=r$ (the number of
 bosons,
 {\it i.e.}, the rank of $\g$). This construction  constitutes a the direct extension of that performed on the circle $\mathbb S^1$, but adapted to the two-dimensional manifols $\mathbb S^1\times
\mathbb S^1$ andr $\mathbb S^2$.\\

 In order to derive the action of the Virasoro algebra on the Kac-Moody algebra (computing $T_\epsilon P_\epsilon$ and $T_\epsilon U_\epsilon$)  for the two-sphere and the two-torus,
  we apply again the Wick theorem and the regularisation procedure (only $T_\epsilon U_\epsilon$ needs to be regularised).  {\rr After the mentioned regularisation, a routine computation leads to} 
\beqa
\begin{array}{lllc}
T(z,x) P^i(w,y) &=&\displaystyle \Big(\frac{zw}{(z-w)^2} P^i(w,y) + \frac{z}{z-w} w \partial_ w P^i(w,y) \Big)\delta(x-y)
\ , \ \  |z| >|w| \nn\\[0.6ex]
T(z,x) U^\alpha(w,y) &=&\displaystyle \Big(\frac{z^2}{(z-w)^2} U^\alpha(w,y) + \frac{z}{z-w} w \partial_ w U^\alpha(w,y)\Big)
\delta(x-y) \ , \ \  |z| >|w|
\end{array}
\eeqa
where $x=\theta, y=\varphi$ for $\mathbb S^1 \times \mathbb S^1$ and $x=u, y=v$ for $\mathbb S^2$.  After extraction of the modes we arrive at the expressions
\beqa
\label{eq:VirKM}
\begin{array}{llllc}
\big[L_{m,p}, H^i_{n,q}\big]&=& -n H^i_{m+n,p+q} &\multirow{2}{*}{for}&\multirow{2}{*}{$\mathbb S^1 \times \mathbb S^1$}\\
\big[L_{m,p}, E_{\alpha, n,q}\big]&=&-n E_{\alpha, m+n,p+q}\\ \\
\big[L_{\ell_1,m_1}, H^i_{\ell_2,m_2}\big]&=& -m_2 c_{\ell_1,m_1,\ell_2,m_2}{}^{\ell_3,m_1+m_2} H^i_{\ell_3,m_1+m_2}
&\multirow{2}{*}{for}
&\multirow{2}{*}{$\mathbb S^2$}\\
\big[L_{\ell_1,m_1}, E_{\alpha, \ell_2,m_2}\big]&=&
-m_2  c_{\ell_1,m_1,\ell_2,m_2}{}^{\ell_3,m_1+m_2} E_{\alpha, \ell_3,m_1+m_2}
\end{array}
\eeqa
 It follows that \eqref{eq:KM-S1xS1} (resp. \eqref{eq:KM-S2}) and \eqref{eq:LL}, \eqref{eq:VirKM} correspond to a bosonic realisation
 of the semi-direct product  of the Kac-Moody  and Virasoro algebras of the  two-torus and the two-sphere, that we denote
 Vir$(\mathbb S^1\times \mathbb S^1)\ltimes \hat \g (\mathbb S^1\times \mathbb S^1)$ and
 Vir$(\mathbb S^2)\ltimes \hat \g (\mathbb S^2)$, respectively.
These two algebras
 can be seen as a direct generalisation of the usual Kac-Moody  and Virasoro algebras  Vir$\ltimes \hat \g$.
The case of the two-torus and the two-sphere are structurally different, \
 as only for the case of the two-torus the embedding Vir$\ltimes \hat \g \subset$ Vir$(\mathbb S^1\times \mathbb S^1)\ltimes \hat \g (\mathbb S^1\times \mathbb S^1)$ holds, {\it i.e.}, we have the semi-direct product of the usual Virasoro and Kac-Moody algebras as a subalgebra.
This observation is a simple consequence of the geometrical properties of the two-torus and the two-sphere, where only the former
can be written as a Cartesian product of one-spheres. So, taking  the two-torus with one of its  radii tending  to zero
(hence corresponding of some `kind' of compactification),
reproduces  Vir$\ltimes \hat \g$ in this limit. Furthermore, in the same limit, the mode expansions \eqref{eq:QT2} lead to the usual modes in String Theory.

\medskip
 Finally, we would like to comment briefly on our regularisation procedure. This observation follows  from  what we already
mentioned  in the fermionic realisation obtained in \cite{rm}. Indeed, the fact that in the case of the two-torus we obtain
the usual Kac-Moody and Virasoro algebras as a subalgebra of their analogue on the two-torus, {\it but with the correct central charges} ($k=1,c=r$),  
legitimates {\it a posteriori} the regularisation related to {  $\zeta-$regularisation} prescription.

\medskip
The Fock space obtained acting on the vacuum \eqref{eq:vac}  with the creation operators
leads to a unitary representation of
 Vir$(\mathbb S^1\times \mathbb S^1)\ltimes \hat \g (\mathbb S^1\times \mathbb S^1)$ and
 Vir$(\mathbb S^2)\ltimes \hat \g (\mathbb S^2)$.
Finally, observe that the generators $-L_{0,0}$  of Vir$(\mathbb S^1\times \mathbb S^1)$ and  Vir$(\mathbb S^2)$ correspond to the differential operators noted by $d$ in \cite{rmm}.  In addition, in the case of the two-torus, only one differential operator is obtained and only one central charge is non-vanishing. This is in accordance with the property of unitarity of the representations of
$\hat \g(\mathbb S^1 \times \cdots\times \mathbb S^1)$, which implies only one non-vanishing central charge. \\

\medskip

\medskip
    {\bb
      We would like to further comment on our regularisation procedure. The main point to obtain a bosonic realisation of the Kac-Moody and Virasoro algebras associated to the two-torus and the two-sphere is related to the consideration of an infinite number of bosons corresponding to a suitable expansion of  the $r$ usual bosons of the circle  $\mathbb S^1$ on
      an extra-dimension associated to the two-sphere or the two-torus.  (See \eqref{eq:QT2} and \eqref{eq:QS2}). Of course, as we have an infinite number of bosons, new divergences appear. These new divergences {\rr are suitably treated with} our Riemann $\zeta-$regularisation.
As explained in \cite{rm}, it may seem that this regularisation {\rr  leads to contradictions.} 
Consider the case of the {\rr two-sphere}. In this case, the infinite number of bosons is simply related to the mode expansion of the usual bosons of $\mathbb S^1$ $q^i(\varphi), p^i(\varphi), \alpha^i_m(\varphi)$ (see \eqref{eq:QT2}).
Introducing regulated bosons, the $\zeta-$function regularisation treats with care infinite sums related to the modes in the $\varphi-$direction. If now we take the large $r$ limit ({\it i.e.}, consider an infinite number of bosons $Q^i(\theta,\varphi), i=1,\cdots,r$), the central charge of the Virasoro algebra diverges and {\it is not} regularised by our procedure. This means that we {\it cannot} consider the large $r$ limit, and the only infinities which are allowed are related to the mode expansion in the $\varphi-$direction.}

    {\bb
We now focus on the two-torus.      
The Hilbert space is then constructed in an easy and standard manner, acting on the vacuum $\big|0\big>$ with creation operators $\alpha^i_{-mp}, m>0, p\in\mathbb Z$.
Differently, if we consider the boson operators $P^i_m(\varphi)$ and $P^i(z,\varphi)$, {\rr  caution needs to be taken}. 
In particular, if one considers the state $\alpha^i_m(\varphi) \big|0\big>=
      \big|i,m,\varphi\big>, m<0$  or $P^i(z,\varphi) \big|0\big>=\big|i,z,\varphi\big>$, a direct computation shows that {\rr
      $$
      \big<i,m,\varphi\big|j, p,\theta\big> =
    \delta_{ij}  \delta_{mp} \delta(\varphi -\theta),\quad 
      \big<i,z,\varphi\big|j, w,\theta\big> =
    \delta_{ij}  \frac{zw}{(z-w)^2} \delta(\varphi -\theta).
    $$}
 This construction extends naturally to the $n-$particle state, but as it is usual in QFT, the normal ordering prescription and the regularisation procedure have to be applied (in particular for the latter when we have
operators  at the same points). Thus, the normal ordering
prescription (possibly {\rr with} the additional $\zeta-$regularisation) must be considered when computing N--points functions.

{\rr Some comments are in order for }the case of the two-torus.      
The {\rr  requirement} to introduce a $\zeta-$regularisation comes from our normal ordering prescription \eqref{eq:no} and order relation \eqref{eq:ord}. Indeed, the generators of  the Kac-Moody algebra  of the two-torus (a similar analysis holds for the generators of the Virasoro algebra) are parameterised by the roots of $\g$ and two relative numbers. The two latter numbers correspond to the Taylor expansion of the generators of the simple Lie algebra $\g$ along the two loops of the two-torus. Stated differently, the necessity of our regularisation procedure is simply the consequence of the fact that we have to order two couples of numbers $(m,p)$ and
$(n,q)$. 
One may wonder {\rr whether it is} possible to avoid this regularisation procedure. {\rr  Actually,} an alternative construction of the representation theory of the toroidal algebras ({\it i.e.}, the Kac-Moody algebra of the two-torus) and its quantum extension where studied in  \cite{MRT, FJMM} (the second paper dealing {\rr specifically} with the quantum version of the toroidal algebra). In these two papers there is no need to define a regularisation prescription. This {\rr results from} a different presentation of the (quantum) toroidal algebra, as in both cases the presentation of the algebra is defined through the mode expansion of the affine (quantum affine) Lie algebra along the circle (in the second article the authors add  central extensions to the loop affine algebra). In particular, in \cite{MRT} a vertex operator was constructed along these lines. Thus, to the roots $\alpha$ of $\hat \g$  a Heisenberg algebra generated by $\alpha(m)$ with $\alpha$ a root of $\hat g$ and $m \in \mathbb Z \setminus\{0\}$ is introduced:
\beqa
\label{eq:Heis}
\big[\alpha(m), \beta(n)\big] = m \delta_{m+n} \alpha\cdot \beta
  \eeqa
  ({\rr In our case} we can easily introduce such a Heisenberg algebra  by setting $\alpha(m,p)= \alpha_i \alpha^i_{m,p}$ with $\alpha$ a root of
  $\g$ with component  $\alpha_i$) and where $\alpha(0)$ is the analogue  of the $p^i$ in our construction. Then within the
  Heisenberg algebra $\alpha(m), m\ne 0$  a Vertex representation of the toroidal algebra was considered (the operator $e^\alpha$ of their construction is the analogue
  of $\hat c_\alpha$ in our construction). This construction is in fact very close to ours, with two {\rr fundamental} differences. {\rr  In first place, as} the expansion involves only one number (the presentation of the algebra is made from the affine Lie algebra), standard techniques of Conformal Field Theory can be used without $\zeta-$regularisation. Secondly, there is an important difference between simple Lie algebras and affine Lie algebras: for the former the determinant of the Cartan matrix is non-zero, whilst for the latter the corank of the Cartan matrix is one. In particular, for affine Lie algebras we have imaginary (or  null) roots (generally denoted
  $n \delta$, {\rr  with} $n\in \mathbb Z\setminus\{0\})$ such that $\delta\cdot\delta=0$. This means in particular that the Heisenberg algebra \eqref{eq:Heis} is degenerate, 
  and consequently  we have zero-norm states in the representation space. Thus the representation obtained in \cite{MRT} is  non unitary, even if $\hat\g$ is an affine Lie algebra associated to a compact Lie algebra. Let us notice that the operators associated to $n\delta$ are basically related to new central charges (beyond $k=1$) since $\alpha(n \delta), n \in \mathbb Z$ are central (see eq.[\ref{eq:Heis}]) and are not considered in our analysis. An interesting question that {\rr emerges from this context and }deserves  further
  investigation is the {\rr precise} relationship between the two constructions. {\rr It is worthy to be recalled that} the construction of \cite{MRT} has been extended to $n-$tori, together with an extension of the Virasoro algebra {\rr close} to our construction (again having {\rr the roots of $\hat g$ and not those of $\g$ as starting point})\cite{MR}.

Let us finally mention that the infinitely many bosons corresponding to the mode expansion of the bosons
  on the two-torus $P^i(z,\varphi), i=1,\cdots, d$ enable us to construct in a  straightforward way the Hilbert space (some care has to be taken only when we consider bosons operators with an infinite number of modes).
         In this construction, there is no need to proceed to  any normal ordering or regularisation prescription since the action of the bosons modes $\alpha^i_{mp}$ on the Fock space is well defined.
         {\rr In other words},  a normal ordering and  regularisation prescription are needed when one tries to define a Kac-Moody and a Virasoro algebra associated to the two-torus, in particular for the computation of the central charge. Having regulated the algebra, it turns out the Fock space associated to the bosons is naturally a unitary representation of the Kac-Moody and Virasoro algebras of the two-torus. This means that  we have, prior to a bosonic realisation of our algebra,
         a unitary representation. Thus to make contact with the  representation space of \cite{MRT}, one may wonder {\rr whether one can extract the zero-norm states and,
           {\rr after extraction of} these states, the two {\rr resulting} representations spaces coincide. }

    }
    
    \medskip
In this note we have extended in a natural manner the usual vertex  realisation of Kac-Moody
and Virasoro algebras to their analogues associated to the two-torus and the two-sphere. The crucial point
to obtain non-trivial central extensions is the normal ordering prescription, where creation operators are
moved to the left of annihilation operators. This normal ordering prescription has to be supplemented by a
 regularisation procedure to regularise infinite divergent
sums. 
 We finally observe that the proposed construction 
extends naturally to the $n-$torus, and  possibly to more complicate manifolds.  In the case of higher dimensional tori, we have the
 canonical embedding chain
\beqa
\text{Vir}(\mathbb T^n)\ltimes \g(\mathbb T^n) \subset \text{Vir}(\mathbb T^{n-1})\ltimes \g(\mathbb T^{n-1})
\subset \cdots \subset  \text{Vir}(\mathbb S^1\times \mathbb S^1)\ltimes \hat \g (\mathbb S^1\times \mathbb S^1)
\subset \text{Vir}\ltimes \hat{\g} \ . \nn
\eeqa
This series of embeddings  again follows easily from  the geometrical properties of $n-$tori, as $\mathbb T^n =
\mathbb T^{n-1} \times \mathbb S^1$,  and can be  properly understood if we consider successively the limits for the various radii tending to zero. {\rr Besides the apparently cumbersome alternative construction with the $\zeta$-regularisation, the procedure presented in this work suggests a quite generic approach that is potentially of use for constructing bosonic realisations of the semi-direct product of the extension of Kac-Moody and Virasoro algebras on higher-dimensional compact manifolds, and where the usual approaches are of difficult implementation. Work in this direction is currently in progress.  }

\bigskip \noindent \textbf{Acknowledgements.}   The authors thank  P. Sorba, A. Marrani, Ch. Schubert and G. Bossard for valuable suggestions and comments.
{\rr We are grateful to the anonymous reviewer for many helpful comments and for pointing out reference \cite{FJMM}.}
RCS  acknowledges financial support by the research
grant PID2019-106802GB-I00/AEI/10.13039/501100011033 (AEI/ FEDER, UE).

\bibliographystyle{utphys}
\bibliography{ref-bos2}

\end{document}